\documentclass[%
 reprint,
 amsmath,amssymb,
 aps,
]{revtex4-2}

\usepackage{graphicx}
\usepackage{dcolumn}
\usepackage{bm}
\usepackage{soul}
\usepackage{xcolor}

\usepackage[per-mode=reciprocal]{siunitx}
\renewcommand{\vec}[1]{\mathbf #1}

\newcommand{\kap}{\kappa}

\newcommand{\vhi}{\varphi}
\newcommand{\sig}{\sigma}
\newcommand{\veta}{\vartheta}

\newcommand{\x}{\vec r}

\newcommand{\Dr}{D_\text{R}}
\newcommand{\Dt}{D_\text{T}}
\newcommand{\nois}{\boldsymbol\zeta}
\newcommand{\kT}{k_\text{B}T}

\begin{document}

\preprint{APS/123-QED}

\title{Motility-induced clustering of active particles under soft confinement}

\author{Timo Knippenberg$^1$}
\author{Ashreya Jayaram$^2$}
\author{Thomas Speck$^2$}
 \email{thomas.speck@itp4.uni-stuttgart.de}
\author{Clemens Bechinger$^1$}%
 \email{clemens.bechinger@uni-konstanz.de}
\affiliation{%
 $^1$Fachbereich Physik, Universität Konstanz\\
 $^2$Institute for Theoretical Physics IV, University of Stuttgart, 70569 Stuttgart, Germany
 }%

\date{\today}

\begin{abstract}
We investigate the structural and dynamic properties of active Brownian particles (APs) confined within a soft annulus-shaped channel. Depending on the strength of the confinement and the Péclet number, we observe a novel re-entrant behavior that is not present in unconfined systems. Our findings are substantiated by numerical simulations and analytical considerations, revealing that this behavior arises from the strong coupling between the Péclet number and the effective confining dimensionality of the APs. Beyond highlighting the important influence of soft boundaries on APs, our research holds significance for future applications of micro-robotic systems.
\end{abstract}

\maketitle

The spatial confinement of systems typically leads to changes of their physical and chemical properties (melting temperature, band structure, magnetic behavior, etc.) compared to their bulk behavior~\cite{RN50,RN51,RN49}.
This also applies to groups of active, i.e. self-propelled, particles (APs), which constitutes the novel class of active matter~\cite{RN7, RN38}. Systems composed of APs are distinguished by an intricate interplay between their local density and propulsion speed which gives rise to a motility-induced phase separation (MIPS)~\cite{RN34,RN15}. This phenomenon has been observed in a wide range of active systems including synthetic colloidal~\cite{RN45,RN53} and bacterial suspensions~\cite{RN63}, collectives of ants~\cite{RN64}, and has been studied extensively in computer simulations~\cite{RN39,RN54,RN55,RN40}.

Motivated by the fact that the natural environment of many living APs is dominated by geometrical confinements, e.g. porous media (soils)~\cite{RN1} or narrow blood vessels~\cite{RN3}, recent studies have investigated their properties near surfaces~\cite{RN25,RN47}, within channels~\cite{RN52}, and confined to optical traps~\cite{RN11}. These studies demonstrate that spatial confinement has a pronounced influence on the behavior of APs promoting, e.g. the formation of lanes and bands~\cite{RN13}. Such confinement-induced behavior is also
important in view of potential applications of synthetic APs such as embolization~\cite{RN3} or drug delivery~\cite{RN56}. Opposed to extensive work on APs in bulk and near \emph{hard} walls and channels, however, only few studies have investigated the behavior of APs confined by \emph{soft} boundaries where the effective confinement strongly depends on the propulsion velocity $v$. Because MIPS has been demonstrated to be largely suppressed~\cite{RN14,RN62} for one-dimensional (1D) confinements (opposed to higher dimensions~\cite{RN39, RN33, RN16, RN17}), this suggests a non-trivial phase behavior of APs in soft confinements as a function of their propulsion velocity.

In this Letter, we experimentally and numerically study the clustering of APs in a narrow annular channel created by two concentric soft repulsive barriers. By varying the softness of these barriers and the velocity $v$ of the APs, we can systematically tune the effective dimensionality of the system from one (single-file) to two spatial dimensions (2D). Upon increasing the AP propulsion velocity or the softness of the confinement, we find a transition from a homogeneous to an inhomogeneous cluster phase and eventually back to a homogeneous AP distribution within the annulus. Such re-entrant behavior is an unique feature of soft confinements and finds no resemblance in 2D systems where phase separation persists for all activities above the critical point.

In our experiments, we use $N$ active Janus particles composed of silica spheres (diameter $\sigma = \SI{7.8}{\micro\meter}$) coated with a $\SI{60}{\nano\meter}$ thick carbon layer on one hemisphere. Due to gravitational and hydrodynamic forces, their motion is restricted to the lower bottom of the sample cell. When suspended in a critical binary mixture of water and propylene-glycol-n-propyl ether (PnP) ($40 \text{ } \% \text{m}$) and illuminated with a laser beam ($\lambda =\SI{532}{\nano \meter}$), the carbon caps are selectively heated. This leads to an asymmetric demixing of the fluid around the particle ~\cite{RN9,RN22}. As a result, the particles perform an active Brownian motion whose propulsion velocity $v$ is controlled by the laser intensity. In our experiments we use a single laser beam which is permanently scanned over all APs and focused on their carbon cap. In the following, we use the Péclet number $\text{Pe}\equiv 3v/(\Dr\sig)$ to characterize $v$ with the rotational diffusion coefficient $\Dr \approx \SI{0.00125}{\second^{-1}}$, see Supplemental Material (SM)~\cite{sm}.

\begin{figure}[t]
\includegraphics{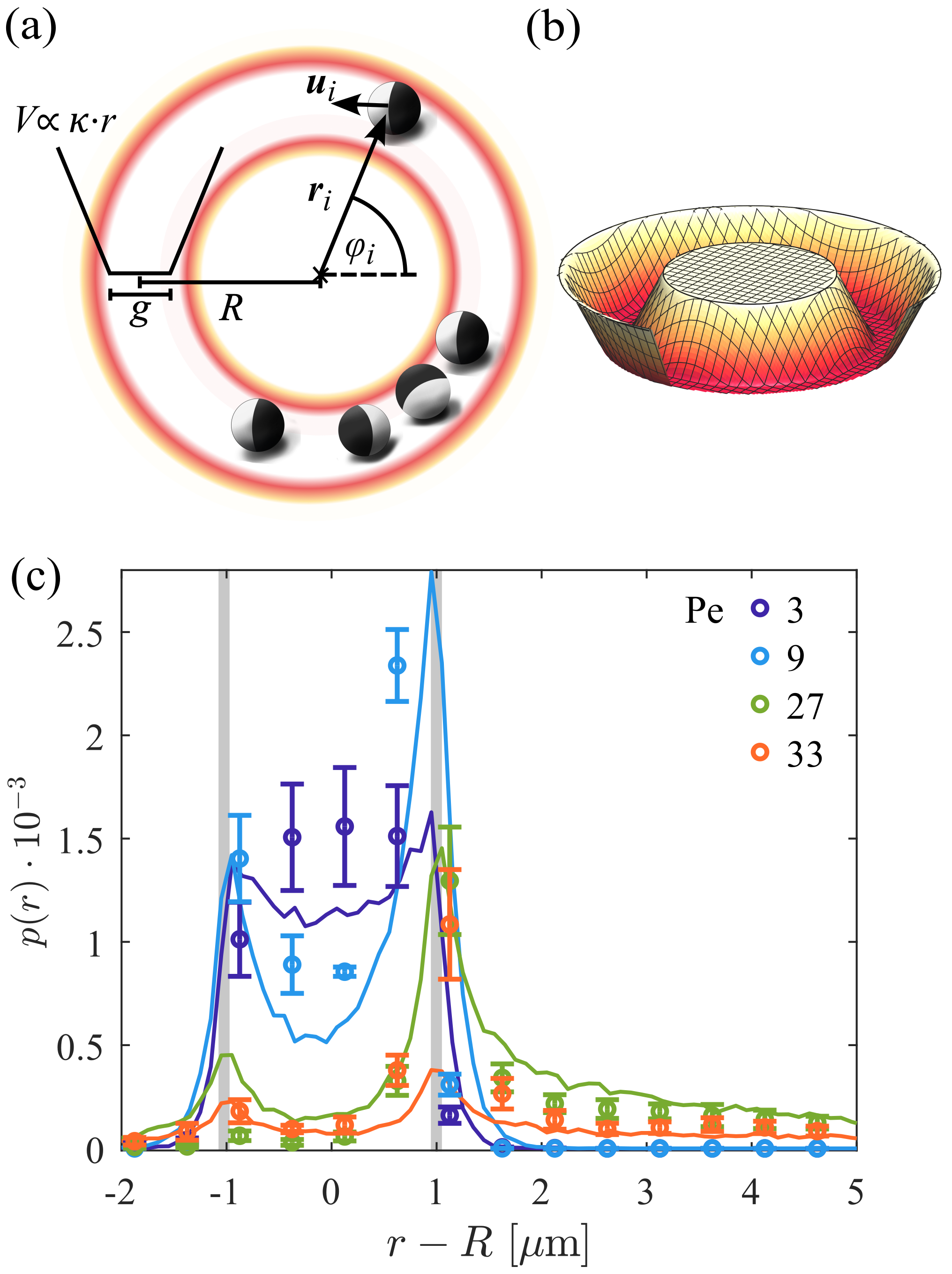}
\caption{\label{fig1} (a) Schematic view of APs confined to an annular-shaped confinement, the latter providing an effective potential as shown in (b). (c) Radial position probability density $p(r)$ of a single AP inside the confinement for different Pe and a confinement strength $\kappa = 70 \frac{k_\text{B}T}{\sigma}$. Circles refer to experimental data, while lines show the corresponding curves from simulations. The gray vertical lines indicate the position of the soft confining walls at $r = R \pm \frac{g}{2}$}
\end{figure}

To create barriers with adjustable softness, we use an additional stationary laser pattern consisting of two concentric rings with mean radius $R = \SI{55}{\micro\meter}$ and gap width $g = \SI{2}{\micro\meter}$ (see in Fig.~\ref{fig1}(a)). When APs approach these rings, the intensity gradient modifies the shape of the demixing regions around APs and thus re-orientates the direction of AP propulsion relative to their orientation~\cite{RN8}. This repulsion from an increasing light gradient leads to an effective external potential $V(r)$ (Fig.~\ref{fig1}(b)) acting on the APs. The confining strength is controlled by the intensity of the concentric laser pattern~\cite{RN10, RN35}. In contrast to a topographical confinement, this approach avoids the influence of hydrodynamic interactions with the boundaries. 

To experimentally demonstrate the creation of adjustable soft confinements, we determined the radial density $p(r)$ of a single AP as a function of $\text{Pe}$ and the confinement strength $\kappa$, the latter being quantified further below (Fig.~\ref{fig1}(b) and SM). At fixed $\kappa = 70 \frac{k_\text{B}T}{\sigma}$ and $\text{Pe} = 3$, $p(r)$ peaks around $r = R$. For larger Pe the APs are able to increasingly move towards the repulsive barriers which leads to an increasing width of $p(r)$. In addition, two maxima at $r\approx R\pm\SI{1}{\micro\meter}$ appear. They result from the AP accumulation at the inner and outer soft boundary, similar to what is known for their behavior near hard walls~\cite{RN25,RN19,RN20}.
Moreover, as $\text{Pe}$ grows, $p(r)$ becomes highly asymmetric with a tail towards the outer confinement.  
This is due to the preference of APs to accumulate at concave rather than convex boundaries~\cite{RN26,RN27}. A comparable behavior is observed when $\text{Pe}$ is fixed and $\kappa$ is varied (see SM). 

In addition to experiments, we also performed numerical simulations where APs are modeled as active Brownian particles moving in 2D. The dynamics of the position $\x_i$ and orientation $\veta_i$ of the $i$th AP is governed by
\begin{align}
    \dot\x_i &= v\vec u_i + \frac{\Dt}{\kT} \left(-\nabla V_{\text{WCA}} + \vec f_i \right) + \sqrt{2\Dt}\nois_{\text{T},i} \\
    \dot{\veta}_i &= M \sum_{j \in \Omega_i} \sin \left[2\left( \vartheta_j-\vartheta_i\right)\right]\Pi(\veta_i,\veta_j) + \sqrt{2\Dr}\zeta_{\text{R},i},    
\end{align}
where unit orientation vector $\vec{u}_i~\equiv~(\cos\veta_i,\sin\veta_i)^T$ and $\Dt$ is the translational diffusion constant. The components of $\nois_{\text{T},i}$ and $\zeta_{\text{R},i}$ are obtained from a unit normal distribution. The APs interact via the repulsive short-range Weeks-Chandler-Anderson (WCA) potential $V_{\text{WCA}}$~\cite{RN44} modeling volume exclusion. Going beyond the standard model of active Brownian particles, we anticipate lubrication effects to play a major role in a strongly confined system. Previous works on squirmers~\cite{RN29,RN30} and Quincke rollers~\cite{RN32} indicate that these interactions prompt an effective alignment between APs. To account for this, we additionally impose an aligning torque of strength $M$. We find that a choice of $M=10^4\Dr$ results in good agreement between simulations and experiments as discussed below. The indicator function $\Pi(\veta_i,\veta_j)$ is unity if the relative orientation of the two APs is between $\pi/2$ and $3\pi/2$ and zero otherwise. The total torque on the $i$th AP is then the sum over all pairwise torques exerted due to neighboring particles within a circular region $\Omega$ of radius $2^{1/6}\sigma$ centered at $\x_i$. We model the confining force on the $i$th AP as
\begin{equation}
    \vec f_i = 
    \begin{cases}
        \kap\vec e_{r,i} & \text{if } |\x_i|<(R-\frac{g}{2}) \\
        -\kap\vec e_{r,i} & \text{if } |\x_i|>(R+\frac{g}{2}) \\
        0 & \text{otherwise}
    \end{cases}
\end{equation}
with $\vec e_{r,i}\equiv(\cos\vhi_i,\sin\vhi_i)^T$ the normal vector. To validate the assumption of a constant force to emulate the confinement, we calculate the density distribution $p(r)$ of a single AP along the radial direction obtained from simulations. As shown in Fig.~\ref{fig1}(c), the distributions agree well with experiments. At large $\kappa$, however, we observe deviations due to the propulsive force being much weaker than the confining force (for further details, see SM~\cite{sm}).

\begin{figure}[b]
\includegraphics{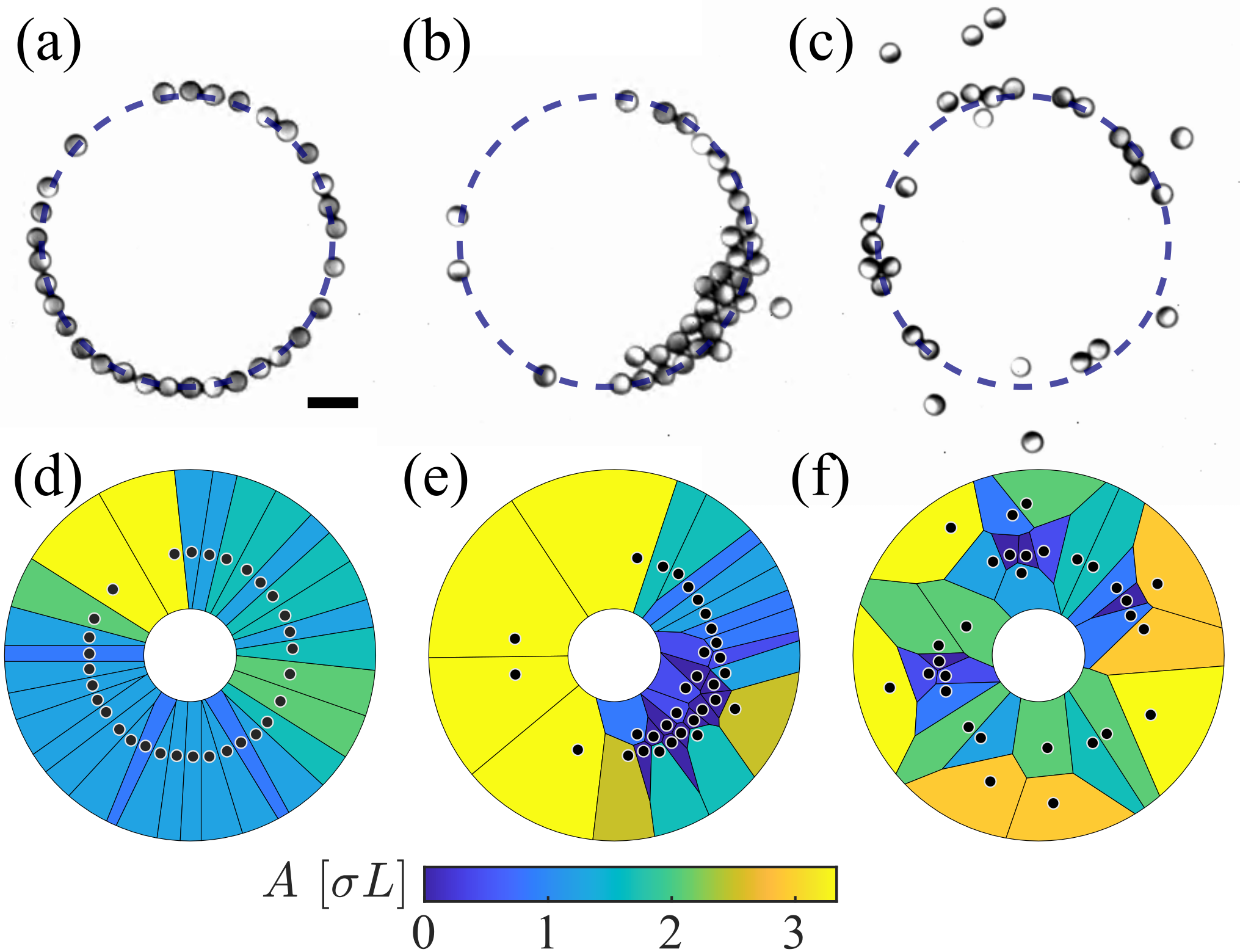}
\caption{\label{fig2} Representative microscopy images for (a) $\text{Pe} = 3$, (b) $27$, (c) $33$. The scale bar represents $\SI{20}{\micro\meter}$. The dashed lines indicates the the channel. (e)-(f) show the corresponding Voronoi tessellations. Colors denote the area $A$ of the Voronoi cells.}
\end{figure}

The interplay of confinement and self-propulsion determines the phase behavior of APs. Representative snapshots of the steady state for $\kappa = 110\frac{\kT}{\sigma}$ and different Pe are shown in Fig. \ref{fig2}(a)-(c) and are obtained for $N=30$. To quantify the particle density, we introduce the filling fraction $\phi\equiv N\sigma/(2\pi R)$, which yields for our experimental conditions $\phi\simeq0.68$. At $ \text{Pe} = 3$, APs are effectively confined to 1D (single file conditions) with only little radial fluctuations. Particles organize into chains which permanently break up into fragments and reform (supplementary videos 1 and 2). Such behavior is in good agreement with previous studies that confirm the absence of macroscopic phase separation of APs in 1D channels~\cite{RN14}. Upon increasing Pe to $\text{Pe} = 27$, APs are able to pass each other and form multi-layered, dense dynamical clusters which coexist with a surrounding gas phase (supplementary video 3). For the conditions shown here, these clusters are comprised of typically more than half of the APs and they are stable over timescales of several $\Dr^{-1}$). Such behavior is comparable to MIPS in 2D systems. Note, however, the phase separation in our experiments is observed for smaller Pe compared to 2D systems, where at least $\text{Pe} \gtrsim 40$~\cite{RN33}. At the largest Pe achieved in our study ($\text{Pe} = 33$), the radial fluctuations of APs further increase which reduces their effective density (supplementary video 4). Similar to 2D systems where MIPS is absent at low densities, also in our case the clusters disappear and the system becomes more homogeneous in angular direction.

\begin{figure}[t]
\includegraphics{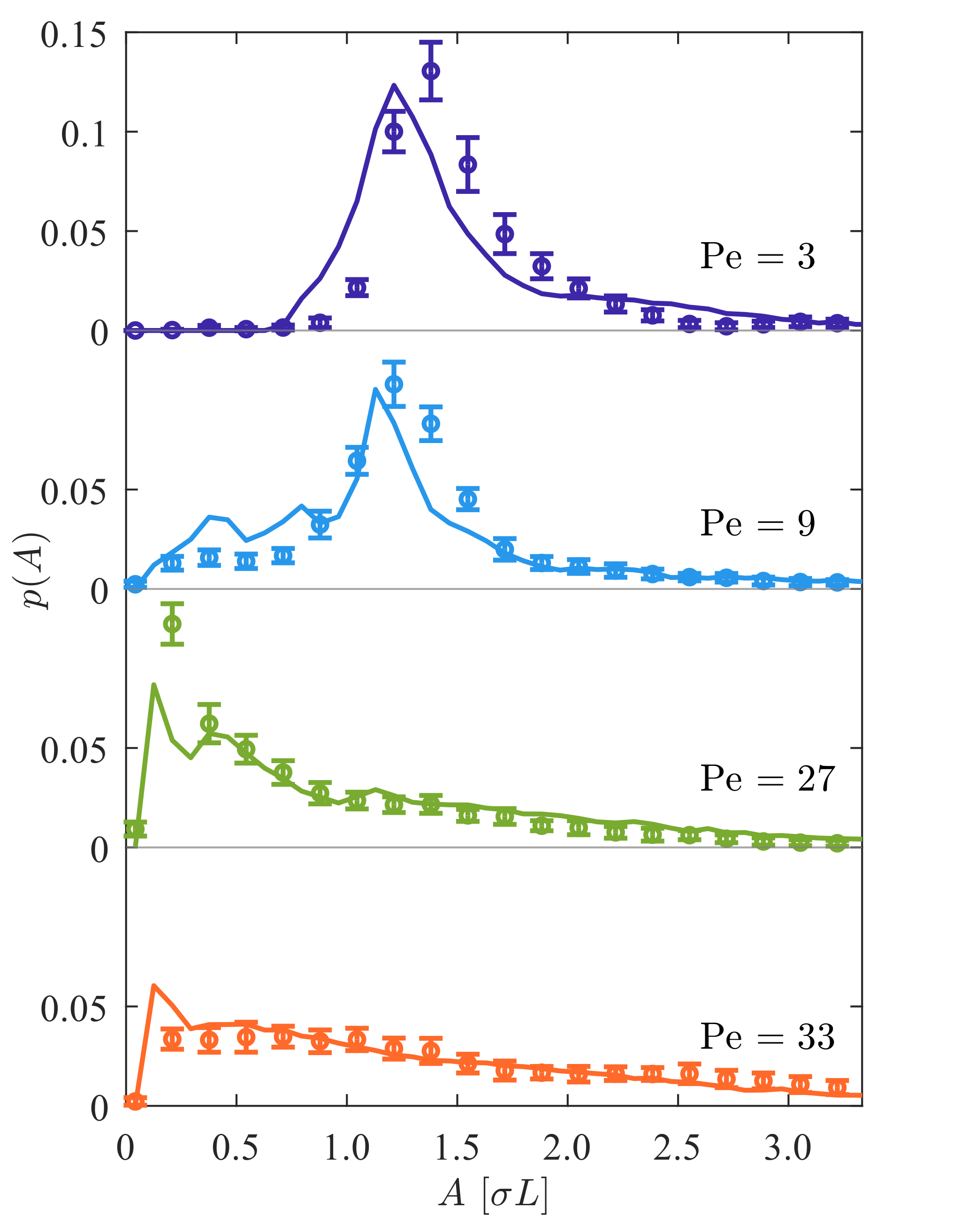}
\caption{\label{fig3} Probability distribution $p(A)$ of Voronoi cell areas for different Pe and fixed $\kappa=110\kT/\sigma, N=30$. Lines represent simulated data, while circles denote experimental data.}
\end{figure}

To quantify the Pe-dependent changes of the AP behavior, we perform a Voronoi tessellation. We restrict the evaluation to an annulus-shaped area whose inner ($R_\text{i}$) and outer ($R_\text{a}$) radii have been chosen to be sufficiently large to include all APs independent of Pe (here, $L = R_\text{a}-R_\text{i} = 9.6 \sigma$). Fig.~\ref{fig2}(d)-(f) show the corresponding Voronoi cells to Fig.~\ref{fig2}(a)-(c) where the Voronoi areas $A$ are colored according to their values. The time-averaged (each measurement was averaged over $\approx 36 \Dr^{-1}$) probability distribution $p(A)$ is shown in Fig.~\ref{fig3}. For $\text{Pe}=3$ and $9$, $p(A)$ is rather narrow which indicates that APs are homogeneously spread in angular direction within the annulus. With increasing $\text{Pe}=27$, this peak gradually disappears at the expense of a maximum which develops at small $A$ exhibiting a long tail towards larger values. Such behavior indicates the presence of an AP cluster. At the largest $\text{Pe}=33$ the maximum shrinks, leading to a rather more uniform $p(A)$. This corresponds to a random distribution \cite{RN42} of APs within the annulus (Fig.~\ref{fig2}(f)).

Comparable results are also obtained from our numerical simulations. Here the confinement strength $\kap$ has been used as a fitting parameter and best agreement to the data previously shown was obtained for $\kap=110 \frac{\kT}{\sigma}$ (solid lines in Fig.~\ref{fig3}). The presence of a lubrication-induced torque ($M \neq 0$) is crucial to facilitate the escape of APs in lateral direction upon direct collisions. This promotes the formation of multi-layer clusters in agreement with our experiments.

\begin{figure}[t]
\includegraphics{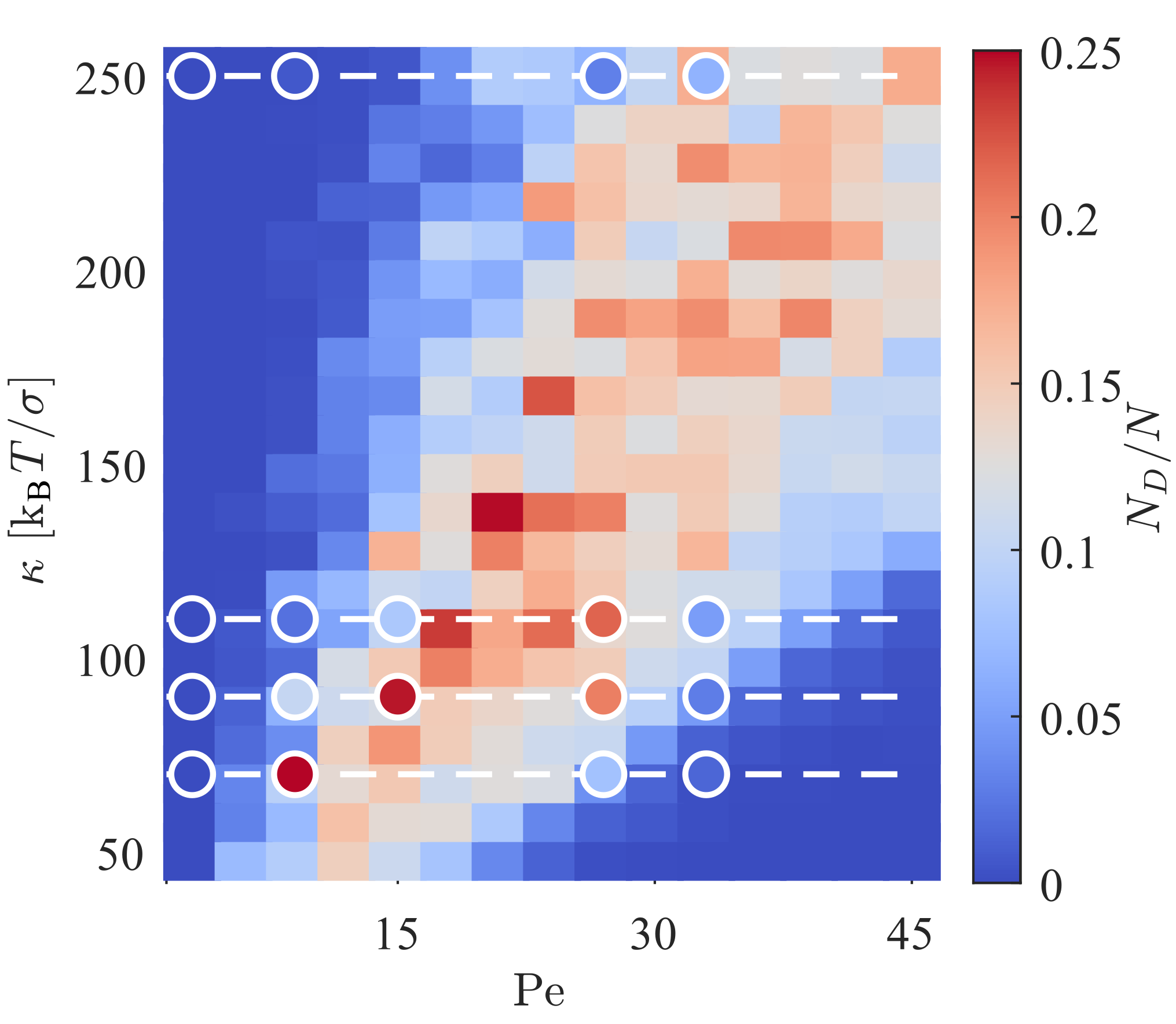}
\caption{\label{fig4} Phase diagram showing the fraction of APs in a dense surrounding $N_D/N$ for different $\kappa$ and $\text{Pe}$. Here $N=30$. The rectangles are simulation data and the circles represent experiments. The dashed lines are guides to the eye.}
\end{figure}

In Fig.~\ref{fig4}, we show the phase diagram of APs in an annular confinement as a function of $\kap$ and Pe. As an order parameter to characterize the presence of clusters, we have measured the fraction $\langle N_\text{D}/N \rangle$ of APs having a Voronoi area below a threshold value~\cite{RN1}. In our case this threshold was set to be a bit larger than the corresponding value for a 2D closest packing $A< 0.2 \sigma L  \approx 2\sig^2$. Clearly, cluster formation is limited to a diagonal region in the $\kap$-Pe space which suggests a re-entrant behavior along both axes. A similar behavior was also found for $N=46$ ($\phi\simeq1.04$), i.e. slightly overcrowded systems (SM~\cite{sm}).

We can rationalize the phase diagram in Fig.~\ref{fig4} from simple geometric considerations. To focus on the essential ingredients, we consider a flat channel confined by perpendicular forces with strength $\kap$. We exploit that for constant force, the density decays exponentially to very good approximation~\cite{RN59,RN60,RN61} with sedimentation length
\begin{equation}
    \xi = \frac{\kT}{\kap}\left(1+\frac{v^2}{2\Dr\Dt}\right).
\end{equation}
This length is reduced as the confining force is increased while APs with larger speed are able to move against the force and to explore a larger area. We model the density distribution as equal to $\rho_0$ within a stripe of width $\sigma$ and to decay as $\rho(y)=\rho_0e^{-|y|/\xi}$ outside. Normalizing the density leads to a dimensionless filling fraction ${ \phi_0=\rho_0\sigma^2=\phi/\left[1+2\xi/\sigma e^{-\sigma/(2\xi)}\right] }$ within the force-free gap.

\begin{figure}[t]
\includegraphics{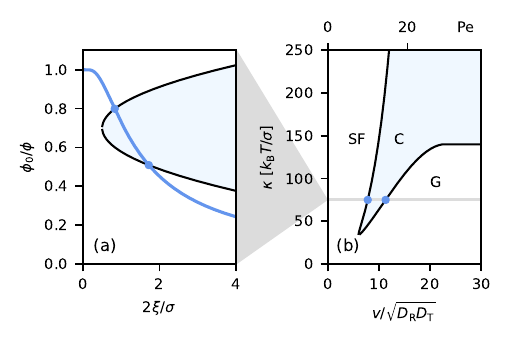}
\caption{\label{fig5} (a)~Filling fraction $\phi_0$ (blue line) within a straight confined channel as a function of sedimentation length $\xi$. Also shown are the coexisting filling fractions $\phi_\pm$ of clusters (black lines) due to the motility-induced instability above the critical speed $v_\text{c}$ for $\kappa=75\kT/\sigma$. (b)~Theoretical phase diagram in the same representation as Fig.~\ref{fig4}. The lines delineate the inhomogeneous cluster regime (C) from single file (SF) behavior at small speeds and large forces and the weakly confined homogeneous gas (G). The top axis shows the P{\'e}clet number estimated using the measured diffusion coefficients (SM~\cite{sm}). Parameters are: $v_\text{c}=6\sqrt{\Dr\Dt}$, $\phi_\text{c}=0.7\phi$, and $\chi=0.02\sqrt{\Dr\Dt}\phi$.}
\end{figure}

Even though in strict 1D only finite clusters form and complete phase separation is preempted~\cite{RN57, RN58}, the dynamical instability underlying motility-induced phase separation persists. To expose the basic mechanism, we employ the mean-field form~\cite{RN43} of the coexisting filling fractions ${ \phi_\pm=\phi_\text{c}\pm\chi\sqrt{v^2-v^2_\text{c}} }$ with critical fraction $\phi_\text{c}$, critical speed $v_\text{c}$, and shape coefficient $\chi$. Figure~\ref{fig5}(a) shows these coexisting filling fractions together with $\phi_0$ as a function of the sedimentation length $\xi$. In this representation, $\phi_0$ is invariant but $\phi_\pm$ moves to the left as the confining force $\kap$ is increased ($\xi_\text{c}\sim1/\kap$). Above $\kap>\kap_\text{min}$ both curves cross, defining a range of speeds for which the system becomes inhomogeneous [cf. Fig.~\ref{fig2}(b)]. Increasing the speed further, APs overcome the confining force and thus effectively reduce $\phi_0$, returning the system to the active gas phase [cf. Fig.~\ref{fig2}(c)]. The resulting phase diagram in Fig.~\ref{fig5}(b) reproduces the salient features of the experiments and simulations, cf. Fig.~\ref{fig4}. In particular, there is a range of confining forces $\kap$ for which it predicts re-entrant behavior as the speed $v$ is increased in agreement with the experiments. This demonstrates that such re-entrance is a unique property of soft confinements but not found for hard confinements where the accessible area is independent of activity~\cite{RN46,RN47,RN48}.

In summary, we have experimentally studied the clustering of self-propelled colloids in a soft annular-shaped confinement. In agreement with numerical simulations of a minimal model we find that cluster formation is only present in a narrow regime of the confinement strength and the P{\'e}clet number, otherwise particles are randomly distributed. This behavior results from a strong coupling of the P{\'e}clet number to the radial AP motion which leads to a Pe-dependent change of the effective dimension of the confinement. At low Pe where the APs move in single-file manner, the 1D confinement prevents clustering. At large Pe the increasing radial AP motion reduces the effective particle density which also suppresses clustering (similar to MIPS in two-dimensional systems). Because APs hold great promise as nano-robotic systems, we expect that our study is important in view of how such systems must be operated within soft confinements which are likely to provide realistic conditions in medical applications~\cite{RN3,RN56}. 

%

%
\end{document}